\documentclass[11pt,showpacs,superscriptaddress,amsmath,amssymb,nofootinbib]{revtex4}

\usepackage{graphicx}
\usepackage{dcolumn}
\usepackage{bm}
\usepackage{amssymb}
\usepackage{amsmath}
\usepackage{epsfig}    
\usepackage{color}
\usepackage{slashed}
\usepackage{float}
\newcolumntype{I}{!{\vrule width 1.3pt}}
\begin{document} 
\title{Enhancements of weak gauge boson scattering processes at the CERN LHC}

\author{Cheng-Wei Chiang}
\email{chengwei@ncu.edu.tw}
\affiliation{Department of Physics and Center for Mathematics and Theoretical Physics, 
National Central University, Chungli, Taiwan 32001, ROC}
\affiliation{Institute of Physics, Academia Sinica, Taipei, Taiwan 11529, ROC}
\affiliation{Physics Division, National Center for Theoretical Sciences, Hsinchu, Taiwan 30013, ROC}
\author{An-Li Kuo}
\email{101222028@cc.ncu.edu.tw}
\affiliation{Department of Physics and Center for Mathematics and Theoretical Physics,
National Central University, Chungli, Taiwan 32001, ROC}
\author{Kei Yagyu}
\email{keiyagyu@ncu.edu.tw}
\affiliation{Department of Physics and Center for Mathematics and Theoretical Physics,
National Central University, Chungli, Taiwan 32001, ROC}

\begin{abstract}

Current CERN Large Hadron Collider data on the 126 GeV standard model-like Higgs boson suggest the possibility of larger Higgs boson couplings with the weak gauge bosons, $g_{hVV}$, than those in the standard model.  
We use the Georgi-Machacek model as an explicit model to realize such a scenario. 
We find that the $g_{hVV}$ couplings can be larger than the standard model value by a factor of about $1.3$ maximally in the parameter region consistent with the current Higgs boson search data and allowed by various other constraints. 
We then show how the modified $g_{hVV}$ couplings lead to enhancements in various weak boson scattering processes. 
This can be clearly observed as excesses in the transverse mass distributions at around 126 GeV and also the mass of heavy Higgs bosons.   


\end{abstract}
\maketitle
\newpage

\section{introduction }

Since the discovery of a new resonance with mass of around 126 GeV, denoted by $h$, at the CERN Large Hadron Collider (LHC), accumulative data have shown that it is consistent with the standard model (SM) Higgs boson in the production rates of several channels and the spin and CP properties~\cite{Higgs_CMS, Higgs_ATLAS}.  However, before concluding the particle as the SM Higgs boson, more detailed checks have to be done because a SM-like Higgs boson can also exist in models with an extended Higgs sector.  In fact, there is no strong reason why the Higgs sector should be minimal as assumed in the SM.  Models with additional isospin singlet, doublet and/or triplet scalar fields are worth considering as well, particularly in the interest of explaining tiny neutrino mass, existence of dark matter, and/or baryon asymmetry of the Universe.  Therefore, determining the structure of the Higgs sector is of paramount importance in probing new physics. 

One basic experimental constraint on the structure of the Higgs sector is the electroweak $\rho$ parameter, which is measured to be close to unity as a result of the custodial symmetry.  Models with a significant deviation of $\rho$ from unity at tree level are therefore disfavored.  It is known that a Higgs sector composed of only isospin singlets with hypercharge $Y=0$, doublets with $Y=1/2$, septets with $Y=2$~\cite{KKY,Hisano}, and so on has $\rho=1$ at tree level (the electromagnetic charge $Q$ is given by $Q=T_3+Y$ with $T_3$ being the third isospin component throughout this paper).\footnote{
In general, Higgs multiplets with $T$ and $Y$ satisfying with $T=\frac{1}{2}(-1+\sqrt{1+12Y^2})$ does not change $\rho=1$ at the tree level.} 
Such a Higgs sector ({\it e.g.}, in two Higgs doublet models) is regarded as natural.  Nevertheless, there are other models with additional assumptions also predicting $\rho=1$ at tree level.  The Georgi-Machacek (GM) model~\cite{GM,Chanowitz} is a well-known example, in which one complex ($Y=1$) and one real ($Y=0$) triplet scalar fields are added to the minimal Higgs sector, with their vacuum expectation values (VEV's) taken to be the same to maintain the custodial $SU(2)_V$ symmetry.
As a result of the symmetry, masses of the Higgs bosons belonging to the same $SU(2)_V$ multiplet (one 5-plet, one 3-plet and one singlet in the model) are the same at tree level.  Discovering these additional Higgs bosons would be clear evidence of the model, and its phenomenology has been studied recently in Refs.~\cite{GVW1,GVW2,Akeroyd,Chiang_Nomura_Tsumura,Chiang_Yagyu_GM,Englert,Englert2}.

If the Higgs mechanism is fully responsible for the breakdown of electroweak symmetry, studying the couplings between $h$ and the weak gauge bosons, collectively denoted by $g_{hVV}$, serves as another approach to deciphering the structure of the Higgs sector.  Throughout the paper, we take $g_{hVV} = g_{hWW} = g_{hZZ}$ because of the approximate custodial symmetry.  These couplings can be directly learned from scattering processes of weak gauge bosons.  If, due to participation of other scalar bosons in the extended Higgs sector, $g_{hVV}$ couplings deviate from their SM values, an immediate consequence is that the scattering of the longitudinal components of weak bosons becomes strong at high energies ~\cite{Bagger:1993zf,Butterworth:2002tt,Chanowitz:2004gk,Ballestrero:2008gf} until heavier active Higgs bosons show up to unitarize the amplitudes~\cite{Cheung:2008zh}.  The possibility for observing such partially enhanced weak boson scatterings at the LHC had been analyzed in the post Higgs discovery era~\cite{Chang:2013aya}.  The result does not look very promising unless one goes for a high integrated luminosity and high invariant mass for the 14-TeV LHC.  In contrast, a more prominent effect due to the nonstandard $g_{hVV}$ couplings should already be observable at lower energies, as we will show in this work.

Although the precision is still poor, current Higgs search data hint at somewhat larger $g_{hVV}$ couplings than the SM expectation from the viewpoint of global fitting.  This is another motivation for us to consider the GM model, where the $g_{hVV}$ couplings can be enhanced from their SM values~\cite{Akeroyd,Chang:2012gn,KKY,Hisano,Englert}, a feature that is not shared by models with a Higgs sector composed of only doublet and singlet fields ({\it e.g.}, two Higgs doublet models).\footnote{A Higgs sector without $\rho\neq 1$ at the tree level such as the simplest Higgs triplet model can also have larger $g_{hVV}$ couplings.  However, their deviation is as small as $\mathcal{O}(0.1)\%$ due to the constraint from the $\rho$ parameter~\cite{KKY,AKKY}.}  Therefore, verifying the enhanced $g_{hVV}$ couplings would be evidence of the GM model or the like.

In this paper, we perform a $\chi^2$ fit to the current Higgs search data at the LHC within the framework of the GM model.  The $g_{hVV}$ couplings are found to be larger than the SM values, with a maximum reaching $\sim 1.3$ at the 68$\%$ confidence level (CL) while still allowed by other constraints: vacuum stability, perturbative unitarity, $Zb\bar{b}$ data, and electroweak precision data.  In this case, masses of the heavier Higgs bosons are preferably at the order of a few hundred GeV.  Taking the enhanced $g_{hVV}$ couplings and including contributions from the heavy Higgs bosons, we study effects on weak boson scattering processes $W^+W^-(ZZ)\to W^+W^-(ZZ)$, $W^\pm Z \to W^\pm Z$, and $W^\pm W^\pm \to W^\pm W^\pm$ at the LHC~\footnote{
Quite recently, the $W^+W^+\to H_5^{++}$ process in the context of searching for Higgs triplets has been discussed in Ref.~\cite{Englert2}.}.  
The corresponding parton-level processes at the LHC are $qQ\to q'Q'\ell^+\ell^-E_T\hspace{-4.3mm}/\hspace{2mm}$, $qQ\to q'Q'\ell^+\ell^+\ell^-E_T\hspace{-4.3mm}/\hspace{2mm}$, and $qQ\to q'Q'\ell^+\ell^+E_T\hspace{-4.3mm}/\hspace{2mm}$, respectively.  The cross sections of these processes are increased by the enhanced $g_{hVV}$ couplings and the mediation of heavy Higgs bosons.  In general, we find excesses in invariant mass and transverse mass distributions.  In the system of $\ell^+\ell^-E_T\hspace{-4.3mm}/\hspace{2mm}$, excess events are seen at and below the $Z$ pole in the leptonic invariant mass distribution.  A Jacobian-like peak with an edge at around 126 GeV is also seen in the transverse mass distributions.  In the system of $\ell^+\ell^+E_T\hspace{-4.3mm}/\hspace{2mm}$, both distributions have a broad peak with an edge at the mass of the doubly charged Higgs boson.  A similar result is also observed in the $\ell^+\ell^+\ell^-E_T\hspace{-4.3mm}/\hspace{2mm}$ system, with the edge indicating the mass of the singly charged Higgs boson though.

The structure of this paper is organized as follows.  We review the GM model in Section~\ref{sec:model}, where the Higgs potential and the Higgs boson mass spectrum are discussed.  
We also derive the couplings of the SM-like Higgs boson with gauge bosons and fermions. 
In Section~\ref{sec:data}, we perform $\chi^2$ fits of the GM model using the current Higgs boson search data at the LHC.  We also take into account other constraints to the parameter space, including vacuum stability, perturbative unitarity, $Zb\bar{b}$ data and electroweak precision data.  
In Section~\ref{sec:VBF}, we discuss different vector boson fusion processes at the LHC, in hope of testing the enhancement in the $g_{hVV}$ couplings. 
Our findings are summarized in Section~\ref{sec:conclusions}.

\section{The model \label{sec:model}}

In the GM model, the Higgs sector comprises the isospin doublet Higgs field, $\phi$ with hypercharge $Y=1/2$, 
and two isospin triplet Higgs fields, $\chi$ with $Y=1$ and $\xi$ with $Y=0$. 
These fields can be expressed in the forms:
\begin{align}
\Phi=\left(
\begin{array}{cc}
\phi^{0*} & \phi^+ \\
-\phi^- & \phi^0
\end{array}\right),\quad 
\Delta=\left(
\begin{array}{ccc}
\chi^{0*} & \xi^+ & \chi^{++} \\
-\chi^- & \xi^0 & \chi^{+} \\
\chi^{--} & -\xi^- & \chi^{0} 
\end{array}\right), \label{eq:Higgs_matrices}
\end{align}
where $\Phi$ and $\Delta$ transform under $SU(2)_L\times SU(2)_R$ as $\Phi\to U_L\Phi U_R^\dagger$ and $\Delta\to U_L\Delta U_R^\dagger$ with $U_{L,R}=\exp(i\theta_{L,R}^aT^a)$ and $T^a$ being the corresponding $SU(2)$ generators\footnote{
The phase convention for all the component scalar fields $\varphi$ is chosen to be $\varphi^* = +\varphi$.}. 
The neutral components in Eq.~(\ref{eq:Higgs_matrices}) can be parametrized as 
\begin{align}
\phi^0&=\frac{1}{\sqrt{2}}(\phi_r+v_\phi+i\phi_i), \quad 
\chi^0=\frac{1}{\sqrt{2}}(\chi_r+i\chi_i)+v_\chi,\quad \xi^0=\xi_r+v_\xi, \label{eq:neutral}
\end{align}
where $v_\phi$, $v_\chi$ and $v_\xi$ are the VEV's for $\phi^0$, $\chi^0$ and $\xi^0$, respectively. 

The most general Higgs potential invariant under the $SU(2)_L\times SU(2)_R\times U(1)_Y$ symmetry in terms of the fields defined in Eq.~(\ref{eq:Higgs_matrices}) is
\begin{align}
V_H&=m_1^2\text{tr}(\Phi^\dagger\Phi)+m_2^2\text{tr}(\Delta^\dagger\Delta)
+\lambda_1[\text{tr}(\Phi^\dagger\Phi)]^2+\lambda_2[\text{tr}(\Delta^\dagger\Delta)]^2
+\lambda_3\text{tr}[(\Delta^\dagger\Delta)^2]\notag\\
&+\lambda_4\text{tr}(\Phi^\dagger\Phi)\text{tr}(\Delta^\dagger\Delta)
+\lambda_5\text{tr}\left(\Phi^\dagger\frac{\tau^a}{2}\Phi\frac{\tau^b}{2}\right)
\text{tr}(\Delta^\dagger t^a\Delta t^b)\notag\\
&+\mu_1\text{tr}\left(\Phi^\dagger \frac{\tau^a}{2}\Phi\frac{\tau^b}{2}\right)(P^\dagger \Delta P)^{ab}
+\mu_2\text{tr}\left(\Delta^\dagger t^a\Delta t^b\right)(P^\dagger \Delta P)^{ab}, \label{eq:pot}
\end{align}
where $\tau^a$ are the Pauli matrices, $t^a$ are the $3\times 3$ matrix representations of the SU(2) generators,  
and 
\begin{align}
P=\left(
\begin{array}{ccc}
-1/\sqrt{2} & i/\sqrt{2} & 0 \\
0 & 0 & 1 \\
1/\sqrt{2} & i/\sqrt{2} & 0
\end{array}\right). 
\end{align}
When we take $v_\chi=v_\xi \equiv v_\Delta$, 
the $SU(2)_L\times SU(2)_R$ symmetry is reduced to the custodial $SU(2)_V$ symmetry. 
In that case, masses of the weak gauge bosons are given by the same form as those in the SM
\begin{align}
m_W^2 = \frac{g^2v^2}{4},\quad m_Z^2=\frac{g^2v^2}{4\cos^2\theta_W}, 
\end{align}
where $v^2\equiv v_\phi^2+8v_\Delta^2=$(246 GeV)$^2$. 
Thus, the electroweak rho parameter $\rho \equiv m_W^2/(m_Z^2\cos^2\theta_W)$ is unity at tree level. 
One loop corrections to the Peskin-Takeuchi $S$, $T$ and $U$ parameters~\cite{Peskin_Takeuchi} 
have been calculated in Refs.~\cite{GVW2,KKY,Englert}, and are taken into account as constraints in the next section. 

Decomposition of the triplet field $\Delta$ into irreducible representations of $SU(2)_V$, ${\bf 3}\otimes {\bf 3}={\bf 5}\oplus{\bf 3}\oplus{\bf 1}$, tells us that the component scalar states can be classified into the quintuplet, triplet and singlet.  These triplet and singlet states can mix with the other triplet and singlet states from the doublet $\Phi$ (${\bf 2}\otimes {\bf 2}={\bf 3}\oplus{\bf 1}$). 
Consequently, the mass eigenstates include 5-plet scalar bosons ($H_5^{\pm\pm},H_5^\pm,H_5^0$), two sets of 3-plet scalar bosons ($H_3^\pm,H_3^0$) and ($G^\pm,G^0$), and two singlets $h$ and $H_1^0$.  
$G^\pm$ and $G^0$ are the Nambu-Goldstone (NG) bosons for the longitudinal components of the $W^\pm$ and $Z$ bosons.
They are related to the weak eigenstates defined in Eqs.~(\ref{eq:Higgs_matrices}) and (\ref{eq:neutral}) 
via the following transformations
\begin{align}
\left(
\begin{array}{c}
\phi_i\\
\chi_i
\end{array}\right)
&=U_{\text{CP-odd}}
\left(
\begin{array}{c}
G^0\\
H_3^0
\end{array}\right),~
\left(
\begin{array}{c}
\phi^\pm\\
\xi^\pm\\
\chi^\pm
\end{array}\right)
=U_\pm
\left(
\begin{array}{c}
G^\pm\\
H_3^\pm\\
H_5^\pm
\end{array}\right),~
\left(
\begin{array}{c}
\phi_r\\
\xi_r\\
\chi_r
\end{array}\right)=U_{\text{CP-even}}
\left(
\begin{array}{c}
h\\
H_1^0\\
H_5^0
\end{array}\right). 
\end{align}
We assume no CP-violating phases in the potential given in Eq.~(\ref{eq:pot}).  Therefore, the CP properties 
of the scalar boson states are well-defined; namely, $h$, $H_1^0$ and $H_5^0$ are 
CP-even Higgs bosons, and $H_3^0$ is a CP-odd Higgs boson. 
The explicit forms of the transformation matrices are
\begin{align}
U_{\text{CP-odd}}&=
\left(
\begin{array}{cc}
c_H & -s_H \\
s_H & c_H
\end{array}\right),~
U_\pm =\left(
\begin{array}{ccc}
0 & 0 & 0\\
0 & \frac{1}{\sqrt{2}} & -\frac{1}{\sqrt{2}} \\
0&\frac{1}{\sqrt{2}} &\frac{1}{\sqrt{2}} 
\end{array}
\right)\left(
\begin{array}{ccc}
c_H & -s_H & 0\\
s_H & c_H & 0\\
0&0&1
\end{array}
\right)
,\notag\\
U_{\text{CP-even}}&=
\left(
\begin{array}{ccc}
1 & 0 &0\\
0 & \frac{1}{\sqrt{3}} & -\sqrt{\frac{2}{3}}\\
0 & \sqrt{\frac{2}{3}} & \frac{1}{\sqrt{3}}
\end{array}\right)
\left(
\begin{array}{ccc}
 c_\alpha & -s_\alpha &0\\
 s_\alpha & c_\alpha &0\\
0 & 0 & 1 
\end{array}\right), 
\end{align}
where $c_H=\cos\theta_H$, $s_H=\sin\theta_H$, $\tan\theta_H=2\sqrt{2}v_\Delta/v_\phi$, $s_\alpha=\sin\alpha$ and $c_\alpha=\cos\alpha$.  The mixing angle $\alpha$ is given via the relation
\begin{align}
\tan2\alpha =\frac{2(M^2)_{12}}{(M^2)_{11}-(M^2)_{22}}, 
\end{align}
where
\begin{align}
(M^2)_{11}&=8c_H^2\lambda_1v^2, \notag\\
(M^2)_{22}&=s_H^2(3\lambda_2+\lambda_3)v^2+c_H^2M_1^2-\frac{1}{2}M_2^2,\notag\\
(M^2)_{12}&=\sqrt{\frac{3}{2}}s_Hc_H[(2\lambda_4+\lambda_5)v^2-M_1^2].
\label{eq:Meven_ele}
\end{align}
with 
\begin{align}
M_1^2&=-\frac{v}{\sqrt{2}s_H}\mu_1,\quad M_2^2=-3\sqrt{2}s_Hv\mu_2. 
\end{align}
The masses for the 5-plet and 3-plet Higgs bosons are given as
\begin{align}
m_{H_5}^2 &= \left(s_H^2\lambda_3 -\frac{3}{2}c_H^2\lambda_5\right)v^2
+c_H^2M_1^2+M_2^2 ,\quad
m_{H_3}^2 = -\frac{1}{2}\lambda_5v^2+M_1^2,\quad 
\end{align}
and those for the CP-even Higgs bosons are
\begin{align}
&m_{h}^2=(M^2)_{11}c_\alpha^2+(M^2)_{22}s_\alpha^2
+2(M^2)_{12}s_\alpha c_\alpha,\notag\\
&m_{H_1^0}^2=(M^2)_{11}s_\alpha^2+(M^2)_{22}c_\alpha^2
-2(M^2)_{12}s_\alpha c_\alpha.  \label{eq:mass}
\end{align}

A comment on the decoupling limit of the GM model is in order here.  When we take the limit of $M_1^2\to \infty$ (or equivalently $s_H\to 0$ for a fixed value of $\mu_1$), the masses of the 5-plet, 3-plet bosons and $H_1^0$ become infinity, and only the other singlet scalar boson $h$ remains at the electroweak scale~\cite{Chiang_Yagyu_GM}.\footnote{
In Refs.~\cite{GVW1,Englert}, 
the trilinear terms $\mu_1$ and $\mu_2$ are dropped from the potential by imposing a discrete $Z_2$ symmetry 
$\Delta\to -\Delta$.  There is no decoupling limit in that case. 
}  
We thus identify $h$ as the SM-like Higgs boson with the mass of 126 GeV.

In the general case with mixing ({\it i.e.}, non-decoupling case), the couplings between $h$ and the SM fermions and weak gauge bosons can deviate from those of the SM Higgs boson. 
%
The ratios of the couplings of $h$ to the weak gauge bosons and fermions to the corresponding SM values are respectively
\begin{align}
c_{hVV}=c_Hc_\alpha +\frac{2}{3}\sqrt{6}s_Hs_\alpha , \quad
c_{hff}=\frac{c_\alpha}{c_H}.
\label{eq:cratio}
\end{align}
The numerical factor $2\sqrt{6}/3$ in $c_{hVV}$ depends on the representation of the additional Higgs fields and makes $c_{hVV} > 1$ possible.  This feature is not shared by multi-doublet models (including cases with additional singlet fields). 

One may think that, unlike in the SM, the modified weak gauge boson couplings for $h$ in the GM model would spoil the cancellation of $\mathcal{O}(E^2)$ dependence in the amplitude of $V_LV_L \to V_LV_L$ scattering at high energies ($V_L$ denotes the longitudinal component of $W$ or $Z$).  In fact, the unitarity of the scattering amplitudes is restored by including the contributions of the extra Higgs bosons.  For example, in the $W_L^+ W_L^-\to W_L^+ W_L^-$ scattering, 
the $s$-channel and $t$-channel diagrams mediated by $h$, $H_1^0$ and $H_5^0$ give the amplitudes
$-ig^2/m_W^2(c_H^2+3s_H^2) E^2+\mathcal{O}(E^0)$ and $+ig^2/(2m_W^2)(c_H^2+3s_H^2)(1-\cos\theta) E^2+\mathcal{O}(E^0)$, respectively, with $\theta$ being the scattering angle. 
In addition, there is a $u$-channel diagram mediated by $H_5^{\pm\pm}$ that gives the amplitude
$+ig^2/m_W^2s_H^2(1+\cos\theta)E^2+\mathcal{O}(E^0)$. 
The total contribution from the Higgs-exchanging diagrams to the $W_L^+ W_L^-\to W_L^+ W_L^-$ process is then the same as that in the SM at $\mathcal{O}(E^2)$, to be cancelled by the pure gauge contributions. 
If the masses of the extra Higgs bosons are heavy, the cancellation would be delayed until the scale of their masses is reached and the scattering strength becomes stronger in the intermediate scale.  Apparently, a clear observation of significant $V_LV_L \to V_LV_L$ scattering phenomena can serve the purpose of probing the extended Higgs sector. 
Unfortunately, the effects such scatterings at LHC are found to be inconspicuous without demanding a large integrated luminosity and going to the high (multi-TeV) invariant mass regime, as studied in Refs.\cite{Cheung:2008zh,Chang:2013aya}.  This is partly because the SM-like Higgs boson, $h$, has largely unitarized the scattering amplitudes and the cross sections are suppressed by the mass of light quarks involved in the initial state.

In this paper, we study the vector boson fusion processes at the LHC due to the modified $hVV$ couplings in the context of the GM model.  We focus on enhancements in the regime of a few hundred GeV that enjoys the advantage of statistics.
In particular, enhancements can show up in certain distributions of the $VV$ system at around and below 126 GeV.
This approach is useful to test the GM model even if the extra Higgs bosons are too heavy to directly probe at the LHC. 

\section{Data fitting \label{sec:data}}

In this section, we discuss the $\chi^2$ fit to the current Higgs boson search data at the LHC. 
The signal strength is defined by 
\begin{align}
\mu_X^{\text{Ref}} \equiv \frac{\sigma_{h}^{\text{Ref}}\times \text{BR}(h\to X)^{\text{Ref}}}{\sigma_{h}^{\text{SM}}\times 
\text{BR}(h\to X)^{\text{SM}}}, 
\end{align}
where $\sigma_{h}^{\text{Ref}}$ [$\sigma_{h}^{\text{SM}}$] and $\text{BR}(h\to X)^{\text{Ref}}$ [$\text{BR}(h\to X)^{\text{SM}}$]
are the reference value [SM prediction] of the Higgs production cross section and 
that of the branching fraction of the $h\to X$ decay, respectively. 
The latest experimental values of $\mu_X^{\text{exp}}$ are given in Refs.~\cite{CMS_Moriond_gamgam,ATLAS_Moriond,CMS_Moriond} for the ATLAS Collaboration and in Ref.~\cite{CMS_Moriond} for the CMS Collaboration, giving the averaged signal strengths~\cite{Chiang_Yagyu_2HDM}\footnote{
The search for the SM Higgs boson in the decay of $Z\gamma$ mode has also been done by the ATLAS~\cite{ATLAS_Zgam} and the CMS~\cite{CMS_Zgam}. 
The observed 95\% CL upper limit for the cross section is $18.2$ times larger than that in the SM prediction
at the ATLAS for the mass of 125 GeV. 
At the CMS, the limits are about $3-31$ times larger than the SM prediction for the mass region between 120 GeV and 150 GeV.
In Refs.~\cite{Carena_Low_Wagner,Chiang_Yagyu_Zgam}, 
the decays of $h\to \gamma\gamma$ and $h\to Z\gamma$ have been calculated in various models with an extended Higgs sector.  It is pointed out that a comparison between the rates of these two modes is important to 
determining the structure of the Higgs sector.}
\begin{align}
&\mu_{\gamma\gamma}^{\text{exp}} =1.22\pm 0.31,\quad 
\mu_{ZZ}^{\text{exp}} =1.21\pm 0.35,\quad 
\mu_{WW}^{\text{exp}} =0.89\pm 0.27,\notag\\
&\mu_{bb}^{\text{exp}} =0.44\pm 0.87,\quad \mu_{\tau\tau}^{\text{exp}} =0.97\pm 0.55.   \label{ave}
\end{align}
For $\mu_{\gamma\gamma}^{\text{exp}}$ of the CMS Collaboration, we take the value based on the MVA method~\cite{CMS_Moriond_gamgam}.
With the input of Eq.~(\ref{ave}), a $\chi^2$ value can be calculated for each reference value as
\begin{align}
\chi^2 = \sum_X \left(\frac{\mu_X^{\text{exp}}-\mu_X^{\text{Ref}}}{\Delta \mu_X^{\text{exp}}}\right)^2, \label{chisq}
\end{align}
where $\Delta \mu_X^{\text{exp}}$ denotes the standard deviation of $\mu_X^{\text{exp}}$. 

At the same time, we consider various constraints to parameters in the GM model, particularly the mixing angles that enter Eq.~(\ref{eq:cratio}).  First, we include the perturbative unitarity and the vacuum stability as the theoretical constraints. 
The unitarity bound in the GM model has been studied in Ref.~\cite{Aoki_Kanemura}, and can be directly applied to our analysis.  We require that the largest eigenvalue of the $S$-wave amplitude matrix for the elastic scatterings of the two scalar boson states $\langle \varphi_3\varphi_4|a_0|\varphi_1\varphi_2\rangle$ be smaller than 1 in absolute value, where $\varphi_i$ denote generically the scalar bosons in the model.  The vacuum stability condition for the Higgs potential to be bounded from below in any direction of the scalar boson space has been given in Ref.~\cite{Chiang_Yagyu_GM}. 
Secondly, the triplet VEV $v_\Delta$ is constrained by the $R_b$ data because it shows up quadratically in one-loop corrections to the $Zb\bar{b}$ vertex in the model~\cite{Haber_Logan,Chiang_Yagyu_GM}.  Using the current data $R_b^{\text{exp}}=0.21629\pm 0.00066$~\cite{PDG}, the upper bound for $v_\Delta$ is about 62 GeV (70 GeV) at the 95\% CL when the mass of the 3-plet Higgs bosons is taken to be 
300 GeV (500 GeV), corresponding to the constraint on the angle $\theta_H<45.5^\circ$ ($\theta_H<53.6^\circ$)~\cite{Chiang_Yagyu_GM}.   
Furthermore, we take into account the constraints from the $S$, $T$ and $U$ parameters. 
The current data of $S$ and $T$ by fixing $U=0$ are given as~\cite{ST_126}
\begin{align}
S=0.05\pm 0.09,\quad  T=0.08\pm0.07, \label{ST_exp}
\end{align}
where the correlation coefficient is $+0.91$, and the reference value of the SM Higgs boson mass is set at 126 GeV. 
In the GM model, one can tune a counter term in the $T$ parameter 
to fit the data in Eq.~(\ref{ST_exp}). 
Therefore, the $S$ parameter is used to constrain the parameter space.

In subsequent numerical calculations, we take $m_{H_5}=m_{H_3}=m_{H_1}=M_1$ and $M_2=s_H M_1$ corresponding to the case with $\lambda_3=\lambda_5=0$ to enlarge the region allowed by the constraints of perturbative unitarity and vacuum stability.  In that case, we obtain the minimal value of $\chi^2$ ($\chi^2_{\text{min}}$) to be $0.932$ and $0.934$ when we take $m_{H_5}=300$ GeV and ($\alpha$, $\theta_H$)=(12.0$^\circ$,3.90$^\circ$) and $m_{H_5}=500$ GeV ($\alpha$, $\theta_H$)=(11.4$^\circ$, 4.20$^\circ$), respectively, while $\chi^2_{\text{min}}$ is $1.45$ in the SM. 

\begin{figure}[t]
\begin{center}
\includegraphics[width=70mm]{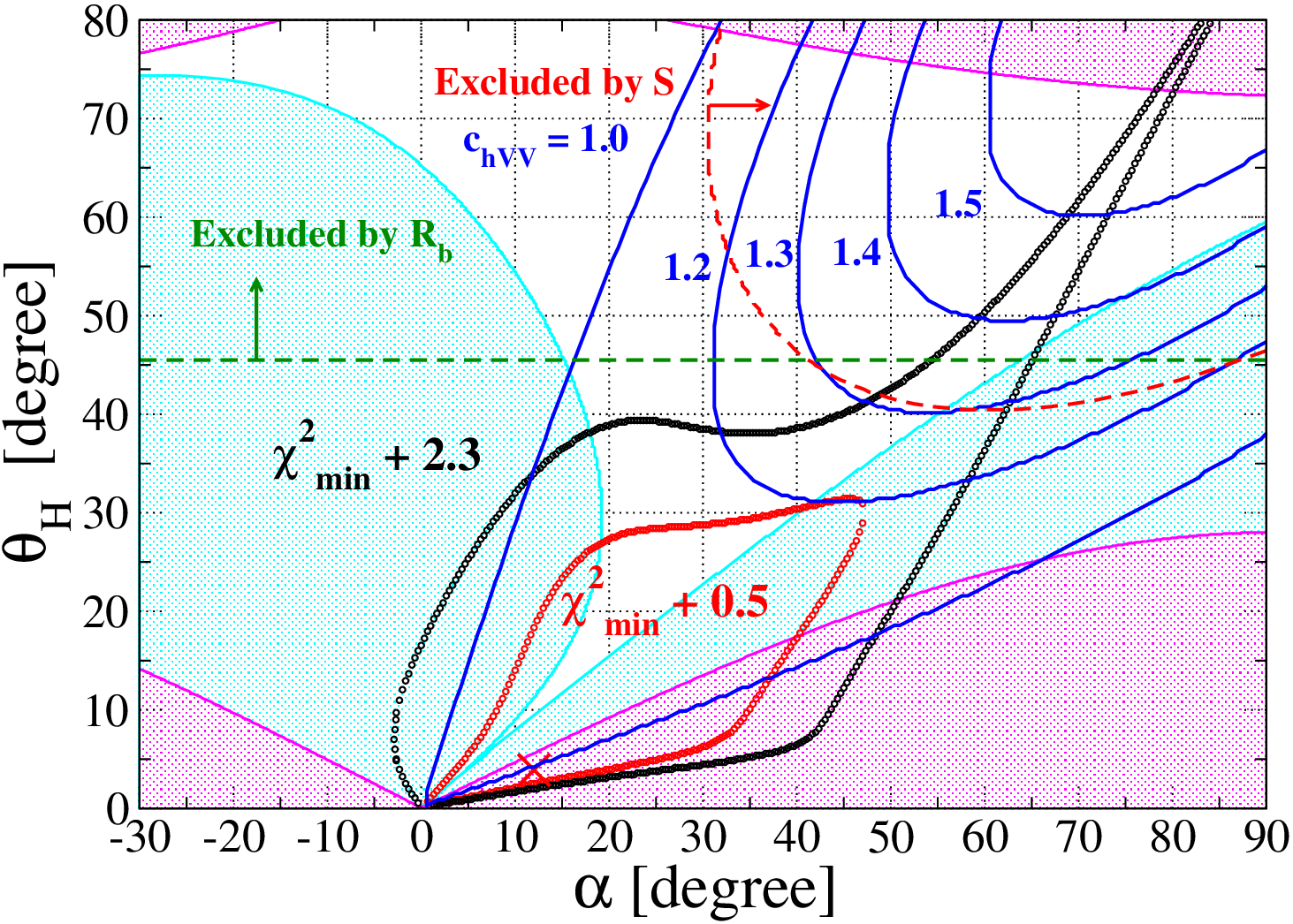}\hspace{3mm}
\includegraphics[width=70mm]{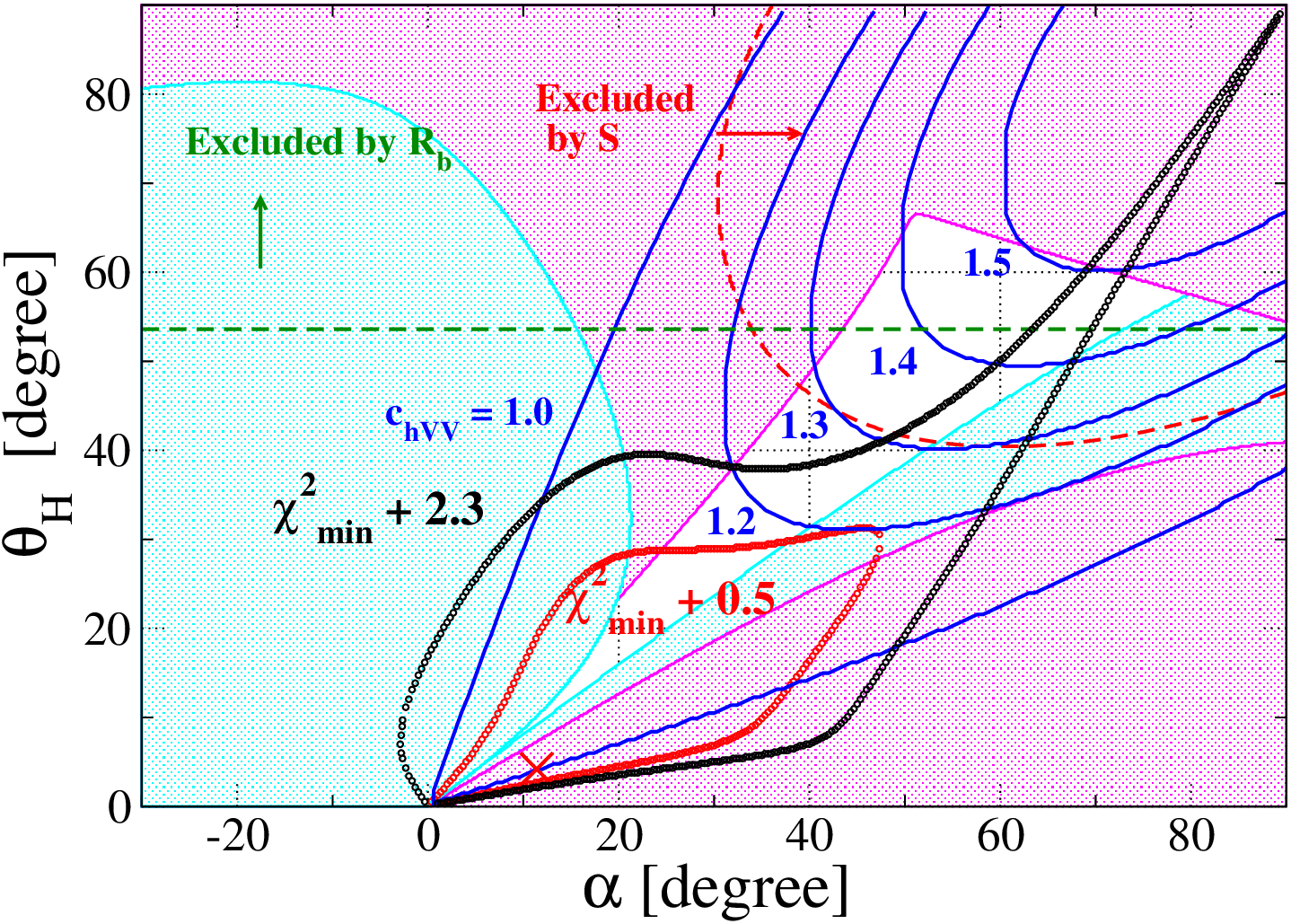}
\end{center}
\caption{
Contour plots of $\chi^2$ values on the $\alpha$-$\theta_H$ plane. 
In the left (right) panel, $m_{H_5}=m_{H_3}=m_{H_1}=M_1$ is taken to be 300 GeV (500 GeV) and 
$M_2=s_H M_1$. 
The red and black curves show respectively the contours of $\chi_{\text{min}}^2+0.5$ and 
$\chi_{\text{min}}^2+2.3$ (corresponds to 1$\sigma$). 
The position of $\chi_{\text{min}}^2$ is marked by a red $\times$. 
The blue curves are the contours of constant $c_{hVV}$.  
The cyan and the magenta regions are excluded by the constraints of vacuum stability and perturbative unitarity, respectively. 
Constraints from the $R_b$ data and $S$ parameter are indicated by the green and red dashed curves, respectively.
}
\label{contour1}
\end{figure}

Fig.~\ref{contour1} shows the contours of $\chi^2$ values on the $\alpha$-$\theta_H$ plane. 
The left (right) panel shows the case with $m_{H_5}=300$ GeV (500 GeV). 
The position of the minimal $\chi^2$ is marked by a red $\times$. 
The black and red curves show respectively the contours of $\chi^2_{\text{min}}+0.5$ and $\chi^2_{\text{min}}+2.3$ (68\% CL).  Each blue curve gives the contour with a constant value of $c_{hVV}$. 
The cyan (magenta) shaded regions are excluded by the vacuum stability (unitarity).  
Constraints from the $R_b$ data and the $S$ parameter are indicated by the green and red dashed curves, respectively, both of which disfavor large values of $\theta_H$.
The contours of $c_{hVV}$ and $\chi^2$ have very little differences between the cases of $m_{H_5}=300$ GeV and $500$ GeV.  However, the region allowed by the vacuum stability and unitarity in the latter case is largely reduced from the former one.  In particular, the change in the unitarity constraint with the change in heavy Higgs boson mass is more significant than that in the vacuum stability.  These two constraints also show that the heavy Higgs boson mass cannot be taken to be arbitrarily high.  It is nevertheless interesting to notice that there is an overlap between the area enclosed by the 1-sigma contour of $\chi^2$ and that allowed by the empirical constraints considered here.  Within 68\% CL, the $hVV$ couplings are allowed to be larger than the SM value, reaching $1.3$ in both cases in the region of $\theta_H\simeq 40^{\circ}$ and $\alpha\simeq 50^{\circ}$-$55^{\circ}$, which corresponds to a deviation in the Yukawa couplings of $c_{hff}\simeq 0.75$-0.84.
Although the current $\chi^2$ minimum sits in the excluded region, we note that any of the following changes can alter the situation.  For example, later experiments favor larger $\mu_{VV}$ and/or smaller $\mu_{bb,\tau\tau}$ values, thereby shifting the minimum into the allowed region in Fig.~\ref{contour1}.  Alternatively, one may take a different $M_1$ value from the heavy Higgs mass to relax the unitarity and vacuum stability constraints, rendering the current minimum point inside the allowed region.

\section{Vector boson fusion processes \label{sec:VBF}}

We now focus on vector boson fusion processes at the LHC to test effects of the enhanced $hVV$ couplings. 
The vector boson scattering processes of interest to us at the parton level are
\begin{align}
qQ \to q'Q'  W^+ W^- (ZZ) ,~~
qQ \to q'Q' W^\pm W^\pm,~~\mbox{and }~
qQ \to q'Q' W^\pm Z, \label{vbf2}
\end{align}
with the produced weak gauge bosons decaying leptonically ($W^\pm\to\ell^\pm\nu$ and $Z\to \ell^+\ell^-$ or $\nu \bar\nu$), where $\ell^\pm$ is $e^\pm$ or $\mu^\pm$ and $q$, $Q$, $q'$ and $Q'$ denote light quarks/antiquarks ($u,d,s,c,\bar{u},\bar{d},\bar{s}$ and $\bar{c}$).
The final-state signatures of these processes are respectively
\begin{align}
j j' \ell^+\ell^- E_T\hspace{-4.3mm}/\hspace{2mm},~~
j j' \ell^\pm\ell^\pm E_T\hspace{-4.3mm}/\hspace{2mm},~~\mbox{and }~
j j' \ell^\pm \ell^\pm \ell^\mp E_T\hspace{-4.3mm}/\hspace{2mm},
\label{vbf}
\end{align}
where $j,j'$ refer to jets.
In Eq.~(\ref{vbf2}), the first process can be mediated by the SM-like Higgs boson $h$ in the $s$-channel, 
the second process in both $t$- and $u$-channels, and the third process in the $t$-channel. 

When the $c_{hVV}$ factor is larger than unity, the gauge-gauge-Higgs couplings for 
the 5-plet Higgs bosons $H_5$ (generically referring to $H_5^{\pm\pm},H_5^{\pm}$, and $H_5^{0}$) and the singlet Higgs boson $H_1^0$ can become important due to the sizeable $v_\Delta$. 
These couplings are given by 
\begin{align}
&(H_5^{\pm\pm}W^\mp W^\mp): +\frac{g}{\sqrt{2}}m_Ws_H,\quad 
(H_5^{\pm}W^\mp Z): -gm_Zs_H,\notag\\
&(H_5^{0}W^+W^-): -\frac{g}{\sqrt{3}}m_Ws_H,\quad 
(H_5^{0}ZZ): -\frac{g}{\sqrt{3}\cos\theta_W}m_Zs_H,\notag\\
&(H_1^{0}W^+W^-): +gm_W(-s_\alpha c_H+\frac{2\sqrt{6}}{3}c_\alpha s_H),\quad 
(H_1^{0}ZZ): +\frac{g}{\cos\theta_W}m_Z(-s_\alpha c_H+\frac{2\sqrt{6}}{3}c_\alpha s_H). 
\end{align}
On the other hand, the 3-plet Higgs bosons $H_3$ (standing for $H_3^\pm$ and $H_3^0$) do not have such interactions at tree level.  Therefore, in addition to $h$, contributions of $H_5$ and $H_1^0$ with sufficiently light mass to the vector boson fusion processes in Eq.~(\ref{vbf2}) need to be included as well. 

We analyze the scattering processes in Eq.~(\ref{vbf}) within the SM, and the GM model with the following parameter choices:
\begin{align}
&(\text{GM13}):~(\theta_H,\alpha)=(40^\circ,55^\circ),~\text{with}~m_{H_5}=m_{H_3}=m_{H_1}=300~\text{GeV},\notag\\
&(\text{GM15}):~(\theta_H,\alpha )=(60^\circ,70^\circ),~\text{with}~m_{H_5}=m_{H_3}=m_{H_1}=300~\text{GeV}. \label{scenario}
\end{align}
For the SM, we only include irreducible backgrounds.
Here GM13 and GM15 correspond respectively to the cases of $(c_{hVV},c_{hff})=(1.3,0.75)$ and $(1.5,0.68)$. 
Although GM15 is disfavored by the current data (see Fig.~\ref{contour1}), we still consider it to see the new physics effects and in case of future data changes. 
We note that although the mass of $H_3$ is not directly related to the vector boson fusion processes, 
it affects how $H_5$ and $H_1$ decay.  For example, if we consider the scenario of $m_{H_3}<m_{H_5},m_{H_1}$, both $H_5$ and $H_1$ can decay into $H_3$ in association with a weak gauge boson.  To avoid such complications, 
we assume that all the masses of extra Higgs bosons are the same, as in Eq.~(\ref{scenario}).  
In this case, $H_5$ mainly decays into a pair of weak bosons because it does not have Yukawa couplings at tree level.
On the other hand, $H_1^0$ can decay into both gauge boson pair and fermion pair, where the Yukawa interactions are derived from the mixing angle $\alpha$. 
We also note that the magnitudes of $H_5VV$ couplings are larger than those of $H_1^0VV$ couplings by one order of magnitude in the cases given in Eq.~(\ref{scenario}).  Therefore, most of the extra Higgs boson contributions to the vector boson fusion processes come from $H_5$.

In order to calculate cross sections and generate events, we use {\tt MadGraph5}~\cite{MG5} for simulations and {\tt CTEQ6L} for the parton distribution functions. 
We impose the following basic kinematic cuts 
\begin{align}
&p_T^\ell >10~\text{ GeV},\quad p_T^j >20 ~\text{ GeV},\notag\\
&|\eta^\ell| < 2.5 ,\quad |\eta^j| < 5.0, \quad \Delta R^{jj} > 0.4, 
\end{align}
where $p_T^{\ell}$ and $\eta^{\ell}$ ($p_T^{j}$ and $\eta^{j}$) are the transverse momentum and pseudorapidity of a charged lepton $\ell$ (jet), respectively, and
$\Delta R^{jj}$ denotes the distance between the two jets. 
In addition to the basic cuts, we also require forward-jet tagging by imposing a large gap in the pseudorapidities of the two jets 
\begin{align}
\Delta \eta^{jj}\equiv |\eta^{j_1}-\eta^{j_2}| > 3.5
\end{align}
to further isolate events from the vector boson fusion processes.

\begin{center}
\begin{table}[t]
\begin{tabular}{c|ccc|ccc|ccc}\hline \hline
Mode&\multicolumn{3}{c|}{$jj'\ell^+\ell^-E_T\hspace{-4.3mm}/\hspace{2mm}$} & \multicolumn{3}{c|}{$jj'\ell^+\ell^+E_T\hspace{-4.3mm}/\hspace{2mm}$}&
\multicolumn{3}{c}{$jj'\ell^+\ell^+\ell^-E_T\hspace{-4.3mm}/\hspace{2mm}$}  \\\hline
Model& SM & GM13 & GM15 &SM & GM13 & GM15&SM & GM13 & GM15   \\\hline
Basic & 85  & 109 & 135 & 7.2 & 16 & 23 &8.7&10&12\\\cline{2-10}
& (203) & (260) & (322)  &(17)&(39)&(57)&(18)&(22)&(26)\\\hline
$\Delta \eta^{jj}$& 18 & 29 & 42 &1.7 &7.6 &12&2.0&3.0&3.9 \\\cline{2-10}
&(51)&(83)&(116) &(5.4)&(22)&(36)&(5.3)&(7.9)&(10.5) \\
\hline\hline
\end{tabular} 
\caption{Cross sections in units of femtobarn (fb) 
for each mode in the SM, GM13 and GM15.  Numbers without (in) parentheses are for collisions at the energy of 
8 TeV (14 TeV).}
\label{cross}
\end{table}
\end{center}

Table~\ref{cross} lists the cross sections of each channel in Eq.~(\ref{vbf}) for the SM and GM model with 
the scenarios of GM13 and GM15 at each step of the kinematical cuts.  Note that we only show the results for positively charged final states in the latter two channels of Eq.~(\ref{vbf}).  
Suppose we take the definition of significance as $S / \sqrt{S+B}$, where $B$ denotes the number of SM background events and $S$ the difference between the number of events in the GM model and that of the SM background.  Then one notices that the significance decreases after imposing the $\Delta\eta^{jj}$ cut.  However, this is not the case in reality because the cut is introduced to effectively remove reducible backgrounds not included in this analysis~\cite{Bagger:1993zf}.

\begin{figure}[t]
\begin{center}
\includegraphics[angle=0,clip,scale=0.32]{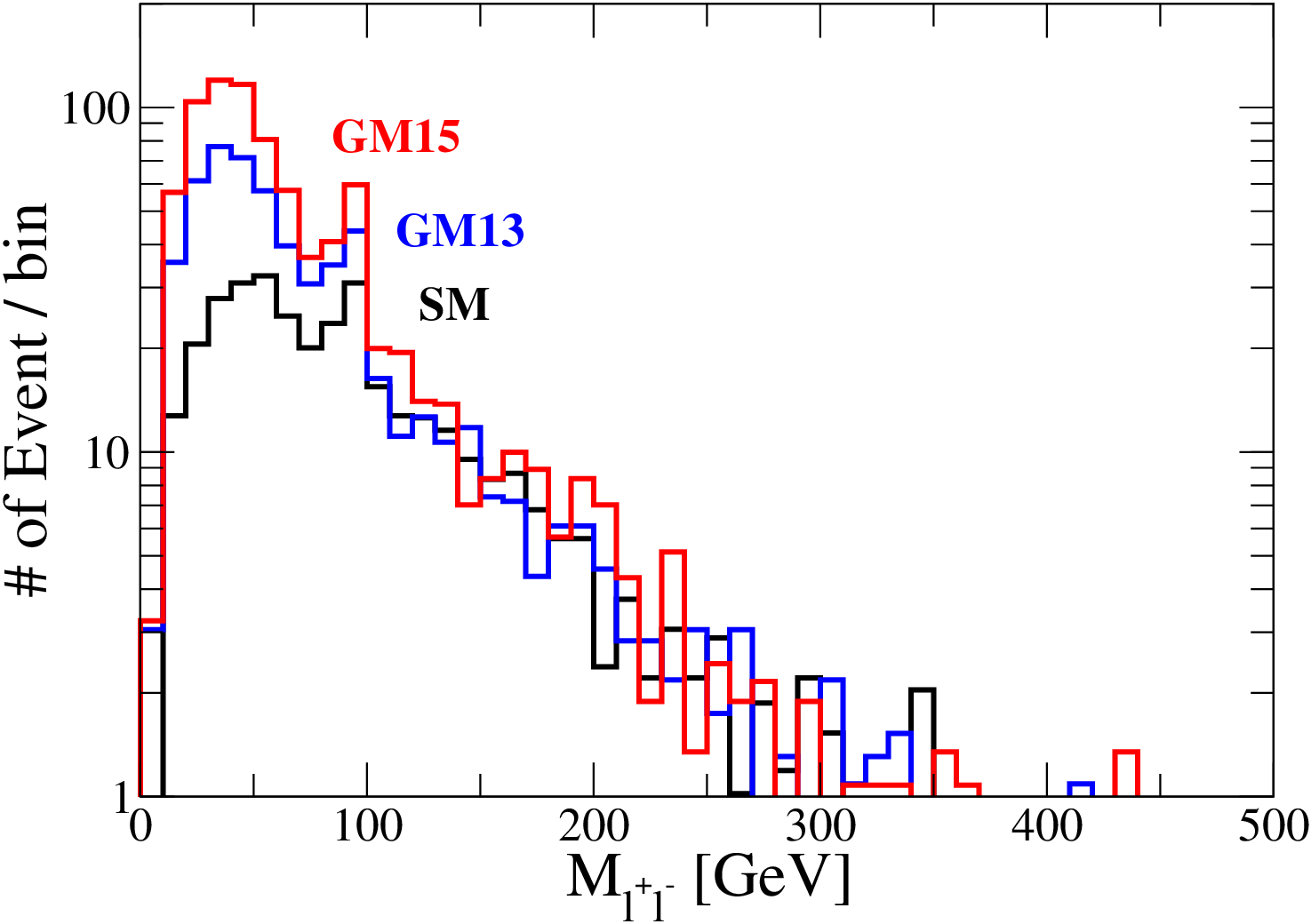}\hspace{3mm}
\includegraphics[angle=0,clip,scale=0.32]{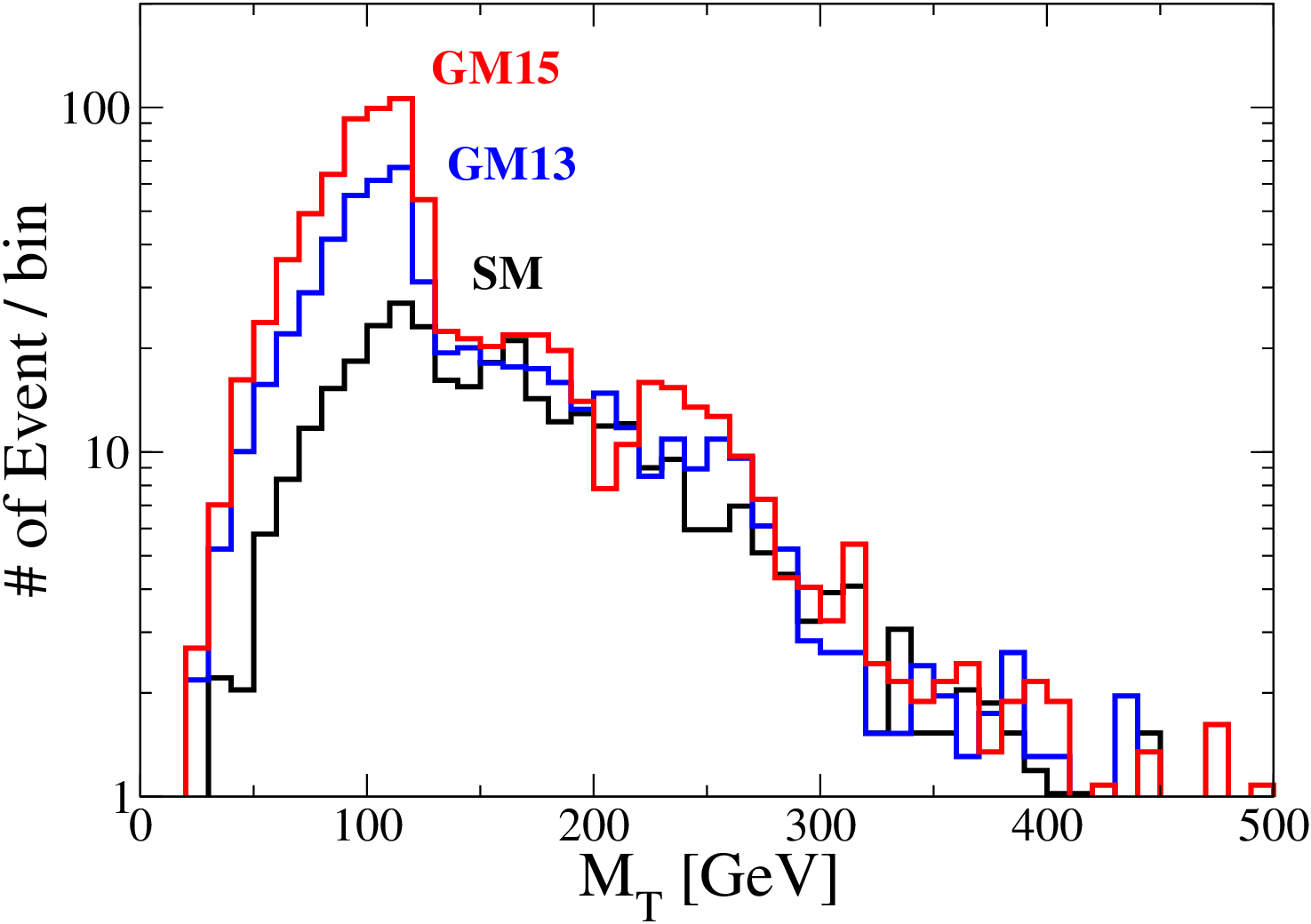}
\caption{Invariant mass distribution of the $\ell^+\ell^-$ system (left panel) and 
transverse mass distribution of the $\ell^+\ell^-E_T\hspace{-4.3mm}/\hspace{2mm}$ system (right panel)
in the $pp\to jj \ell^+\ell^-E_T\hspace{-4.3mm}/\hspace{2mm}$ 
process in the SM (black), GM13 (blue) and GM15 (red) after taking the $\Delta\eta^{jj}$ cut.  The collision energy and the integrated luminosity are 8 TeV and 20 fb$^{-1}$, respectively.}
\label{distr_lplm_8}
\end{center}
\end{figure}

\begin{figure}[t]
\begin{center}
\includegraphics[angle=0,clip,scale=0.3]{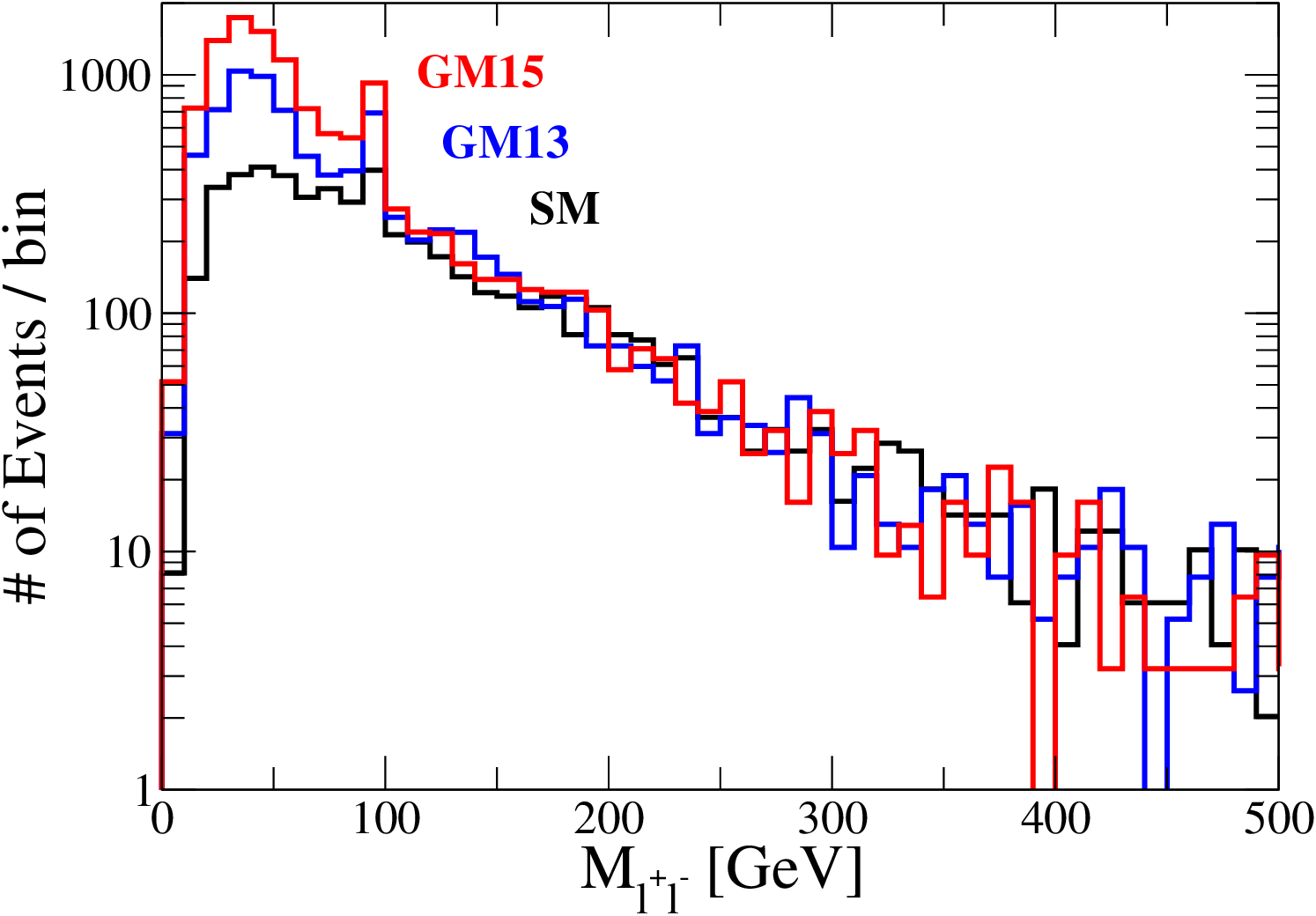}\hspace{3mm}
\includegraphics[angle=0,clip,scale=0.3]{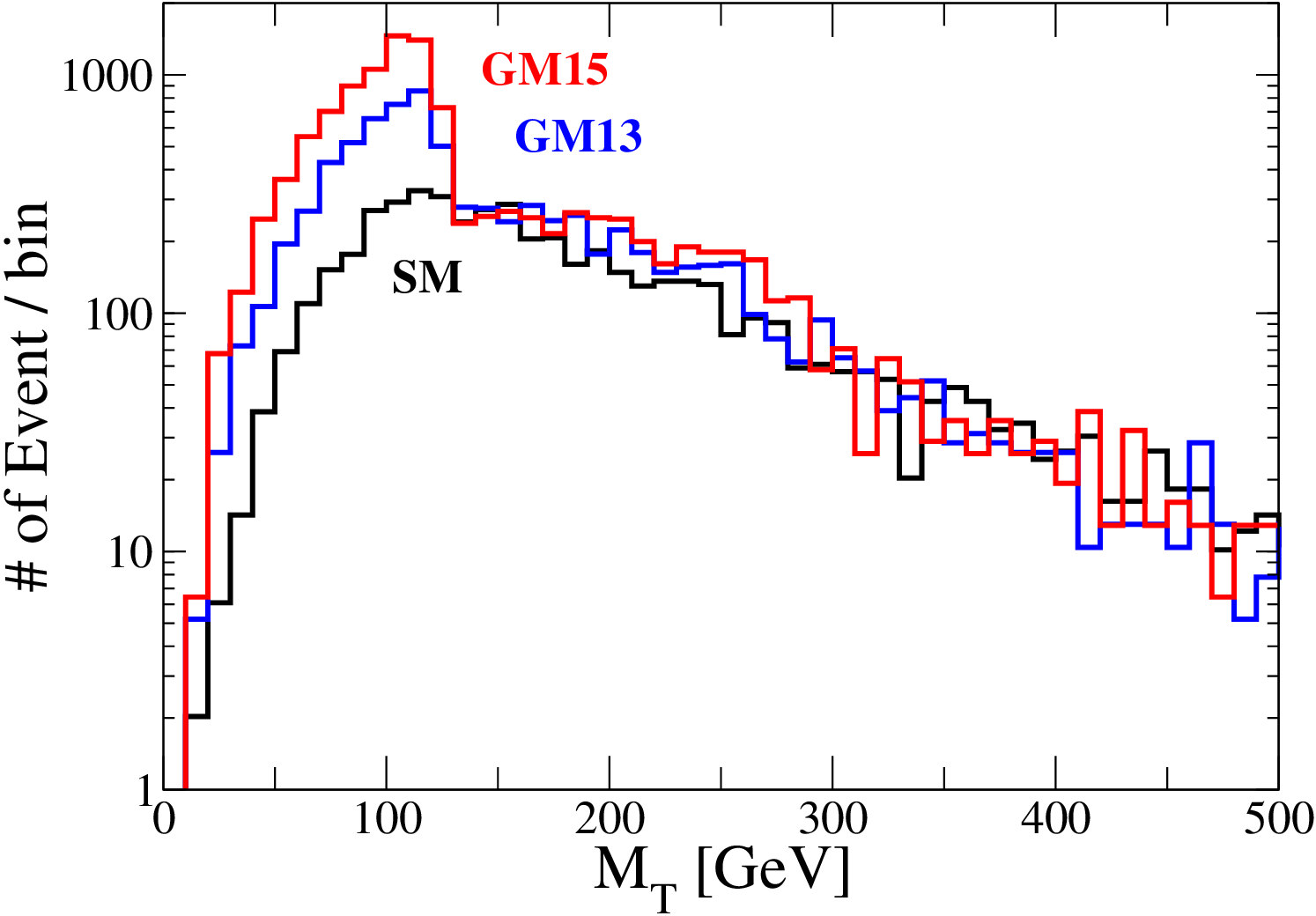}
\caption{Same as Fig.~\ref{distr_lplm_8}, but for the collision energy of 14 TeV and 
an integrated luminosity of 100 fb$^{-1}$. 
} 
\label{distr_lplm_14}
\end{center}
\end{figure}

We consider the invariant mass distribution of the charged lepton system
and transverse mass distribution of the charged leptons with missing transverse momentum system. 
The latter observable is defined by~\cite{transverse_mass}
\begin{align}
M_T^2 &\equiv\left[\sqrt{
M_{\text{vis}}^2+ ({\bm p}_T^{\text{vis}})^2}+| \slashed{\bm p}_T |\right]^2-
\left[{\bm p}_T^{\text{vis}}+ \slashed{\bm p}_T \right]^2,  
\end{align}
where $M_{\text{vis}}$ and ${\bm p}_T^{\text{vis}}$ are the invariant mass and the vector sum of the transverse momenta of the charged leptons, 
respectively, and $\slashed{\bm p}_T$ is the missing transverse momentum determined by the negative sum of visible momenta in the transverse direction.

In Fig.~\ref{distr_lplm_8}, we show the invariant mass distribution of the $\ell^+\ell^-$ system and transverse mass distribution of the $\ell^+\ell^-E_T\hspace{-4.3mm}/\hspace{2mm}$ system in the $pp\to jj\ell^+\ell^-E_T\hspace{-4.3mm}/\hspace{2mm}$ process after imposing the $\Delta\eta^{jj}$ cut.  
The collision energy and the integrated luminosity are taken to be 8 TeV and 20 fb$^{-1}$, respectively.  
In the invariant mass distribution, the number of events in the GM model is greater than that in the SM at around the 40 GeV and 90 GeV, caused respectively by the enhanced $W^+W^-\to h \to W^+W^-$ and $ZZ\to h \to ZZ$ scatterings.  
In the transverse mass distribution, a Jacobian-like peak is seen with the edge at around the Higgs boson mass (126 GeV), more significantly in the GM model than the SM.  A small bump looms in the range of 200 to 300 GeV in the GM13 and GM15 scenarios, as a result of the $H_5$ contribution.

We also show in Fig.~\ref{distr_lplm_14} the same distributions for events simulated under the collision energy of 14 TeV and an integrated luminosity of 100 fb$^{-1}$. 
The shapes of these distributions are almost the same as in Fig.~\ref{distr_lplm_8}, yet the numbers of events are about one order of magnitude larger.

\begin{figure}[t]
\begin{center}
\includegraphics[angle=0,clip,scale=0.32]{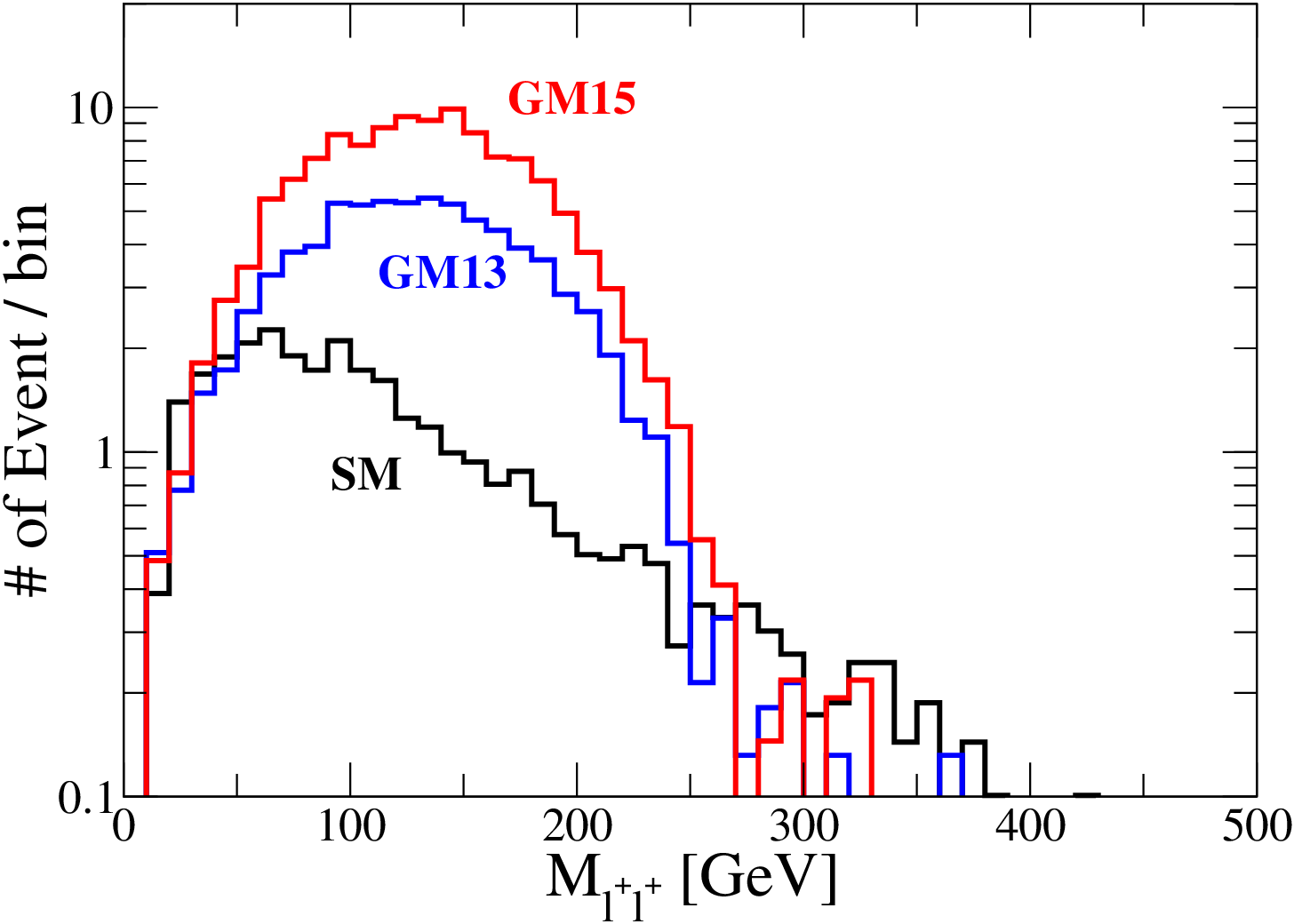}\hspace{3mm}
\includegraphics[angle=0,clip,scale=0.32]{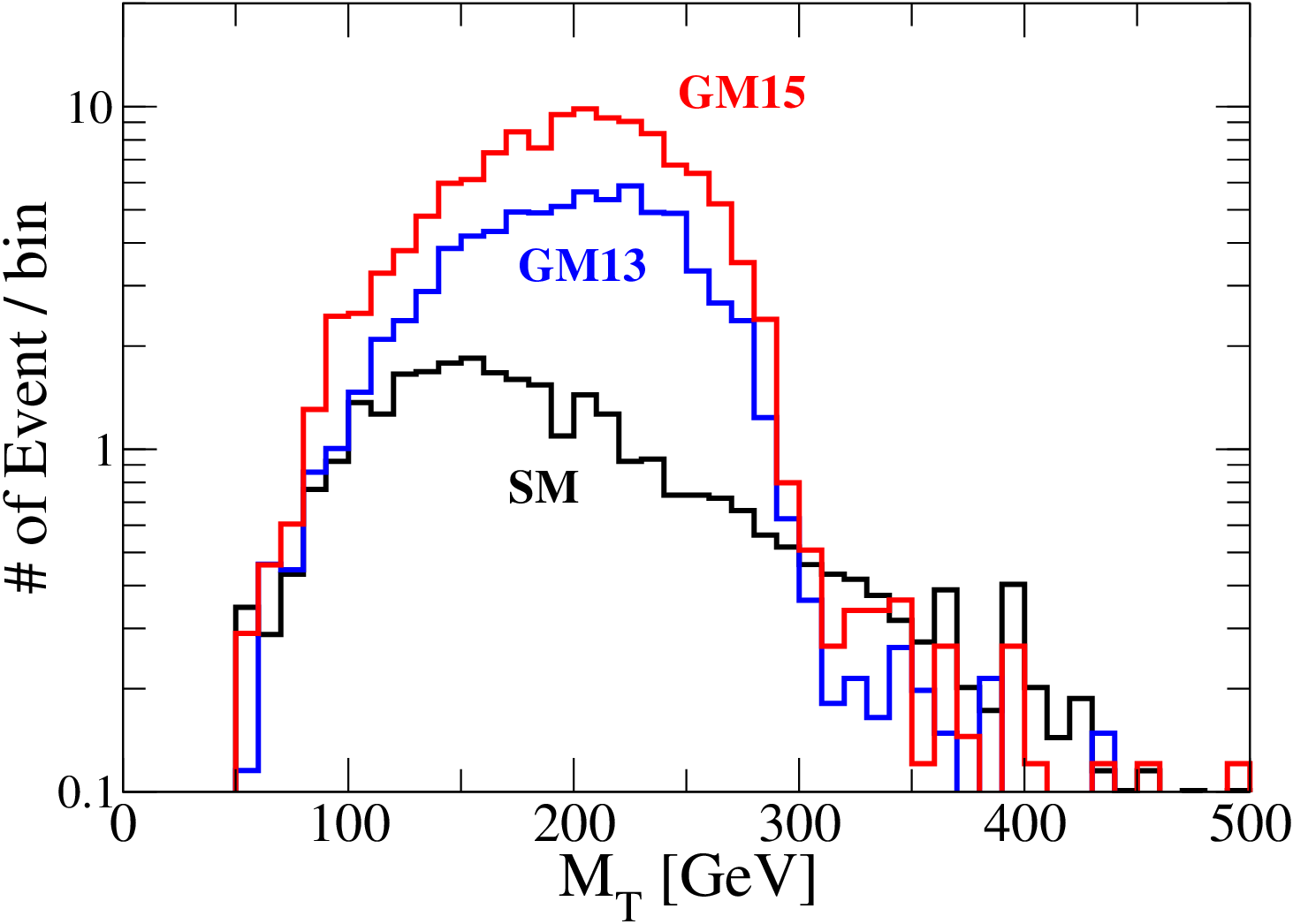}
\caption{Invariant mass distribution of the $\ell^+\ell^+$ system (left panel) and 
transverse mass distribution of the $\ell^+\ell^+E_T\hspace{-4.3mm}/\hspace{2mm}$ system (right panel)
in the $pp\to jj \ell^+\ell^+E_T\hspace{-4.3mm}/\hspace{2mm}$ 
process in the SM (black), GM13 (blue) and GM15 (red) after taking the $\Delta\eta^{jj}$ cut.  The collision energy and the integrated luminosity are 8 TeV and 20 fb$^{-1}$, respectively.}
\label{distr_lplp_8}
\end{center}
\end{figure}

\begin{figure}[t]
\begin{center}
\includegraphics[angle=0,clip,scale=0.32]{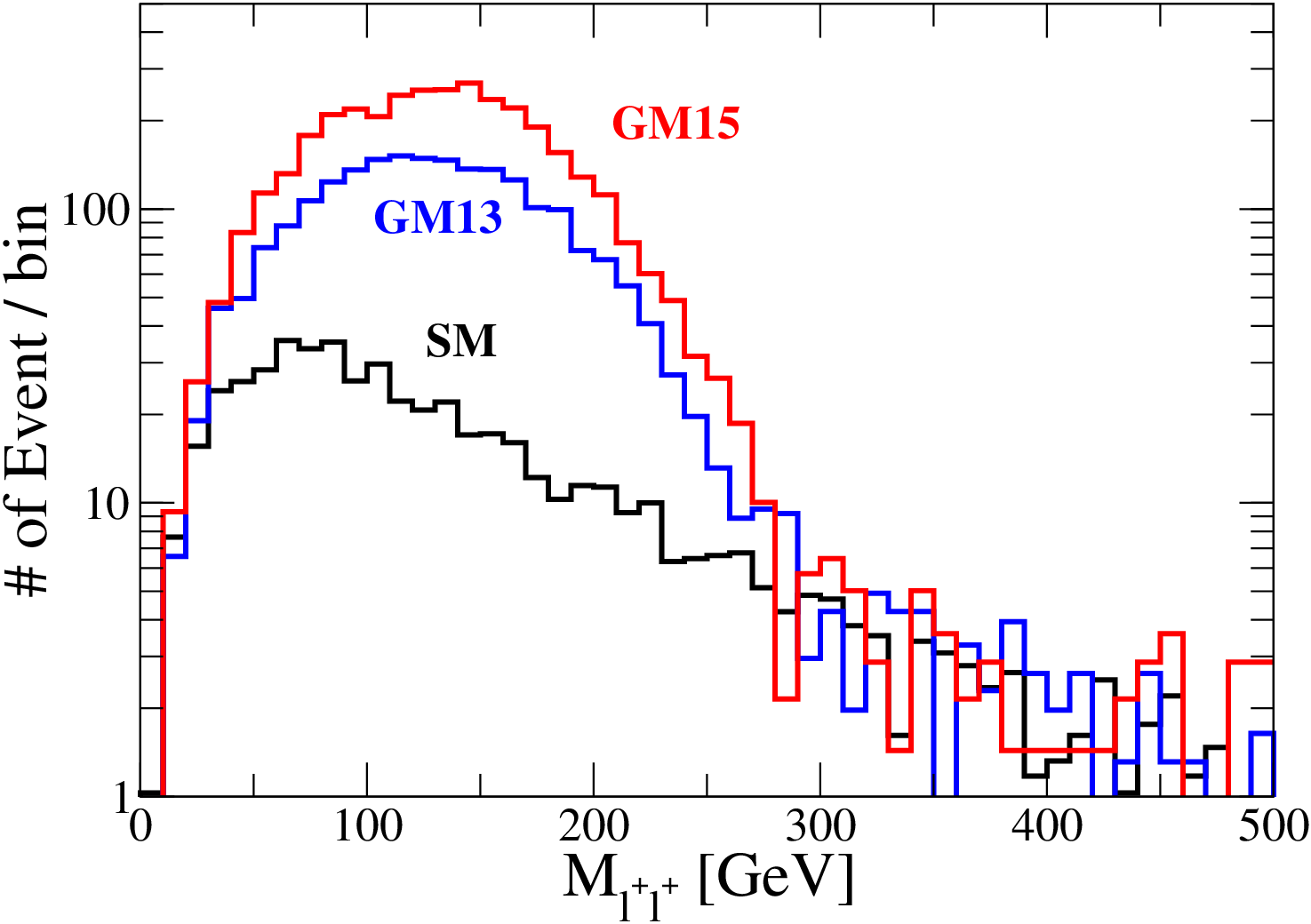}\hspace{3mm}
\includegraphics[angle=0,clip,scale=0.32]{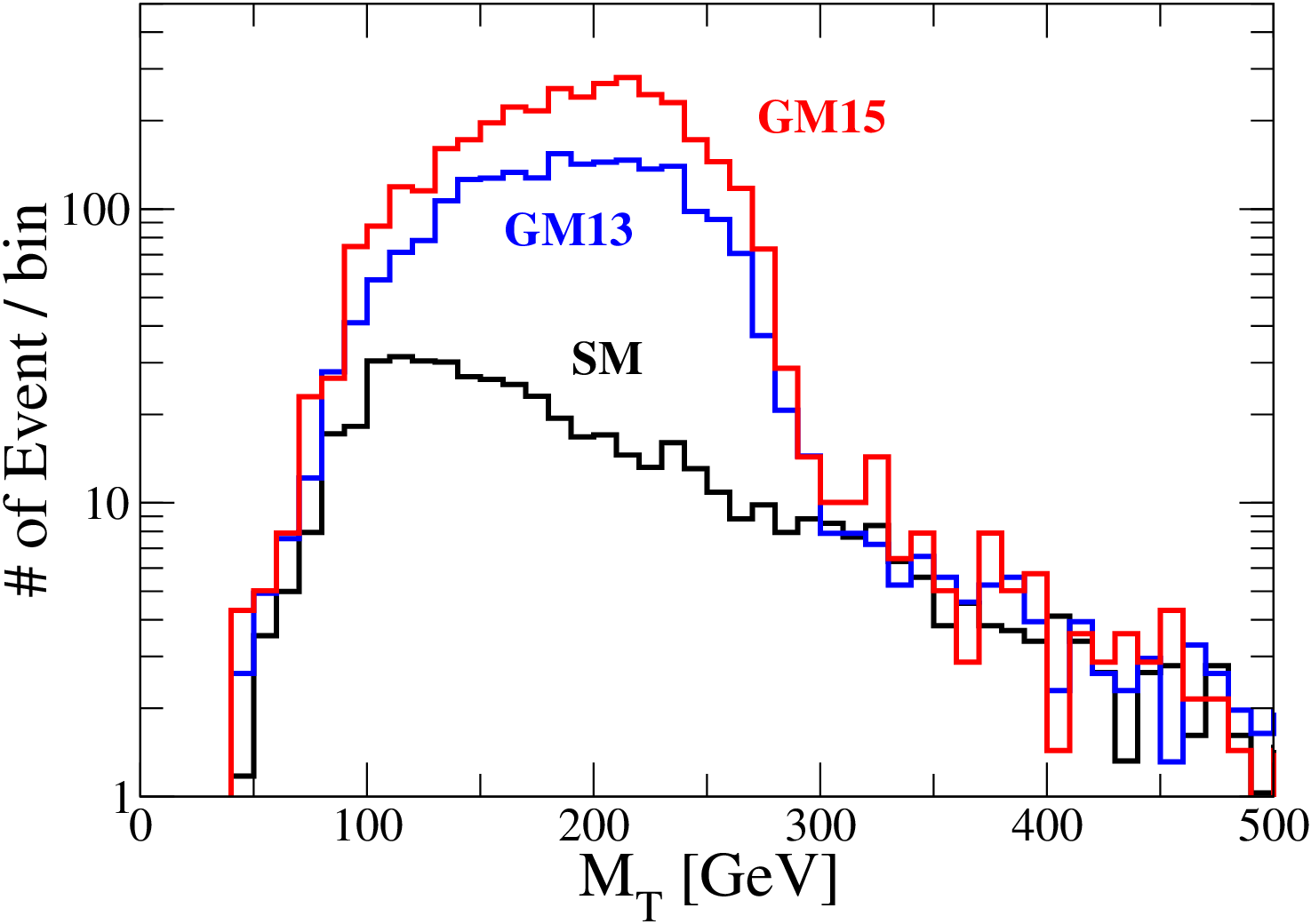}
\caption{Same as Fig.~\ref{distr_lplm_8}, but for the collision energy of 14 TeV and 
an integrated luminosity of 100 fb$^{-1}$.}
\label{distr_lplp_14}
\end{center}
\end{figure}

In Fig.~\ref{distr_lplp_8}, we show the invariant mass distribution of $\ell^+\ell^+$ system and 
the transverse mass distribution of $\ell^+\ell^+E_T\hspace{-4.3mm}/\hspace{2mm}$ system in the 
$pp\to jj\ell^+\ell^+E_T\hspace{-4.3mm}/\hspace{2mm}$ process after imposing the $\Delta\eta^{jj}$ cut. 
For the SM, there is no characteristic feature in both continuum distributions. 
For the GM model, a broad bump peaking at around 150 GeV is seen in the invariant mass distribution.
In the transverse mass distribution, there is a Jacobian-like peak with an edge at around 300 GeV. 
These behaviors can be explained by the $H_5^{++}$ mediation in the $W^+ W^+ \to W^+ W^+$ scattering. 
Similar behaviors, though roughly one order of magnitude larger, can also be observed in the case of 14-TeV collision and 100 fb$^{-1}$ luminosity, as shown in Fig.~\ref{distr_lplp_14}.

\begin{figure}[t]
\begin{center}
\includegraphics[angle=0,clip,scale=0.32]{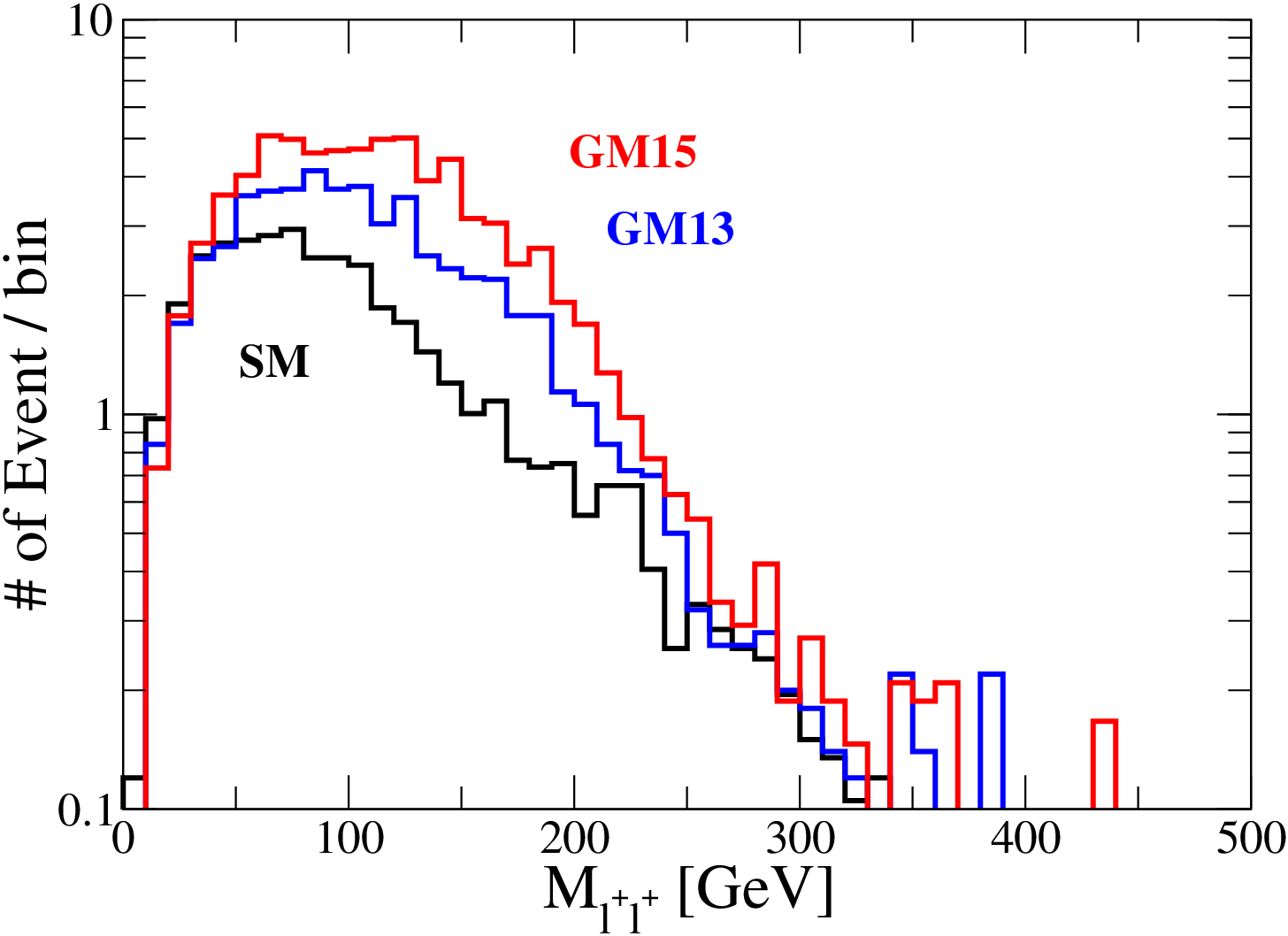}
\hspace{3mm}
\includegraphics[angle=0,clip,scale=0.32]{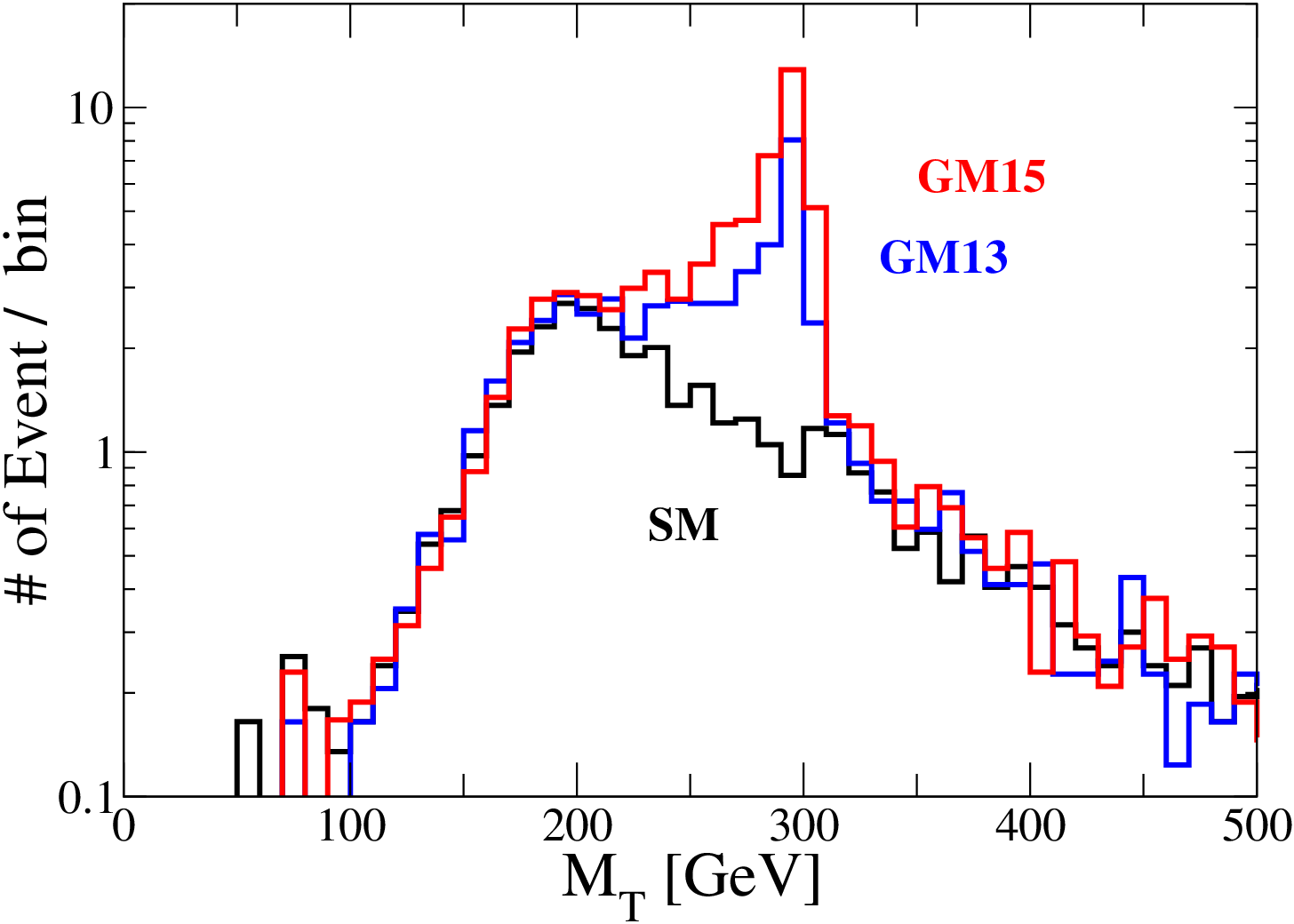}
\caption{The invariant mass distribution of the $\ell^+\ell^+$ system (left panel) 
and transverse mass distribution of the $\ell^+\ell^+\ell^- E_T\hspace{-4.3mm}/\hspace{2mm}$ system (right panel) 
in the $pp\to jj \ell^+\ell^+\ell^-E_T\hspace{-4.3mm}/\hspace{2mm}$ 
process in the SM (black), GM13 (blue) and GM15 (red) after taking the $\Delta\eta^{jj}$ cut. 
The collision energy and the integrated luminosity are 8 TeV and 20 fb$^{-1}$, respectively.}
\label{distr_3l_8}
\end{center}
\end{figure}

\begin{figure}[t]
\begin{center}
\includegraphics[angle=0,clip,scale=0.3]{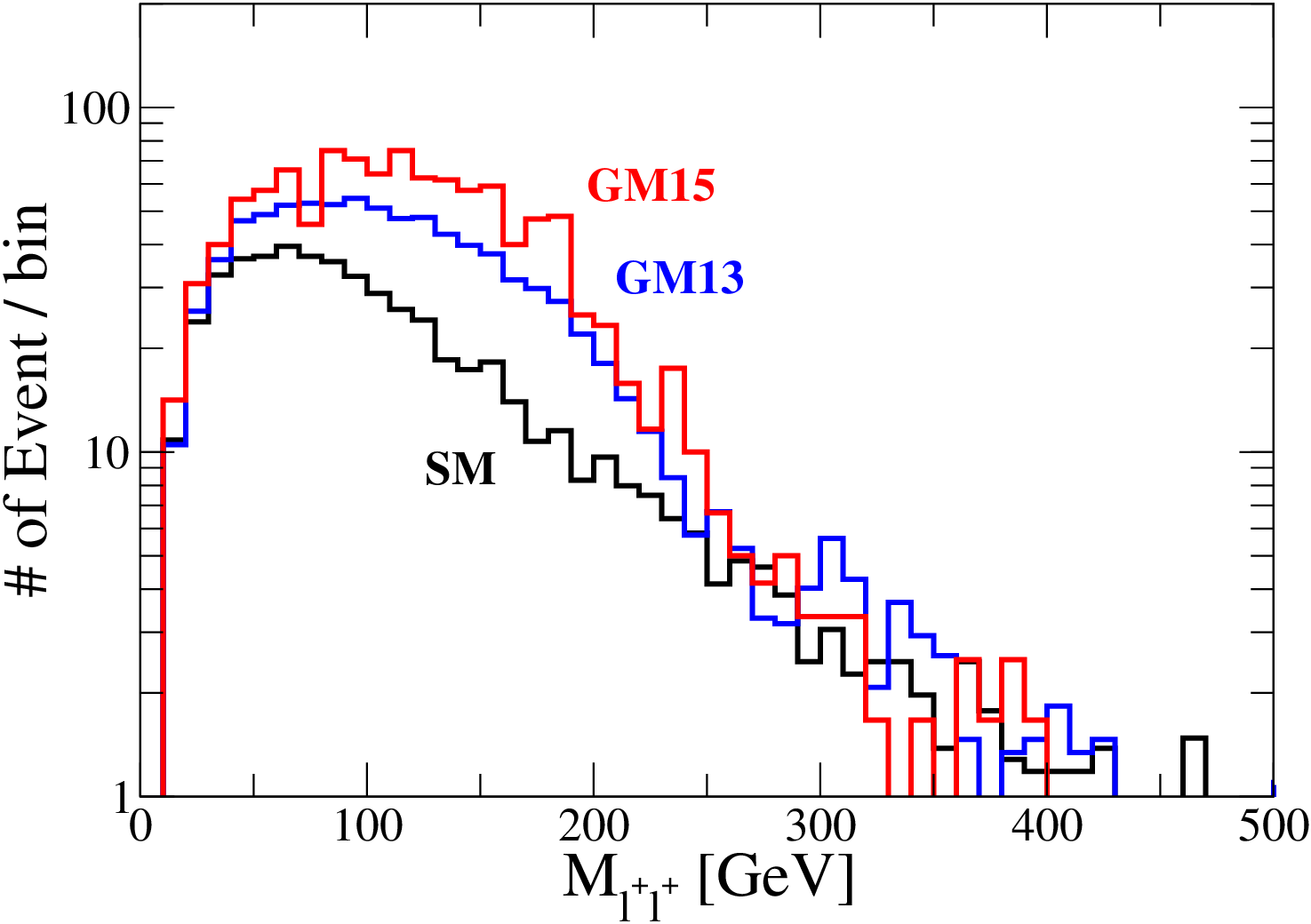}
\hspace{3mm}
\includegraphics[angle=0,clip,scale=0.3]{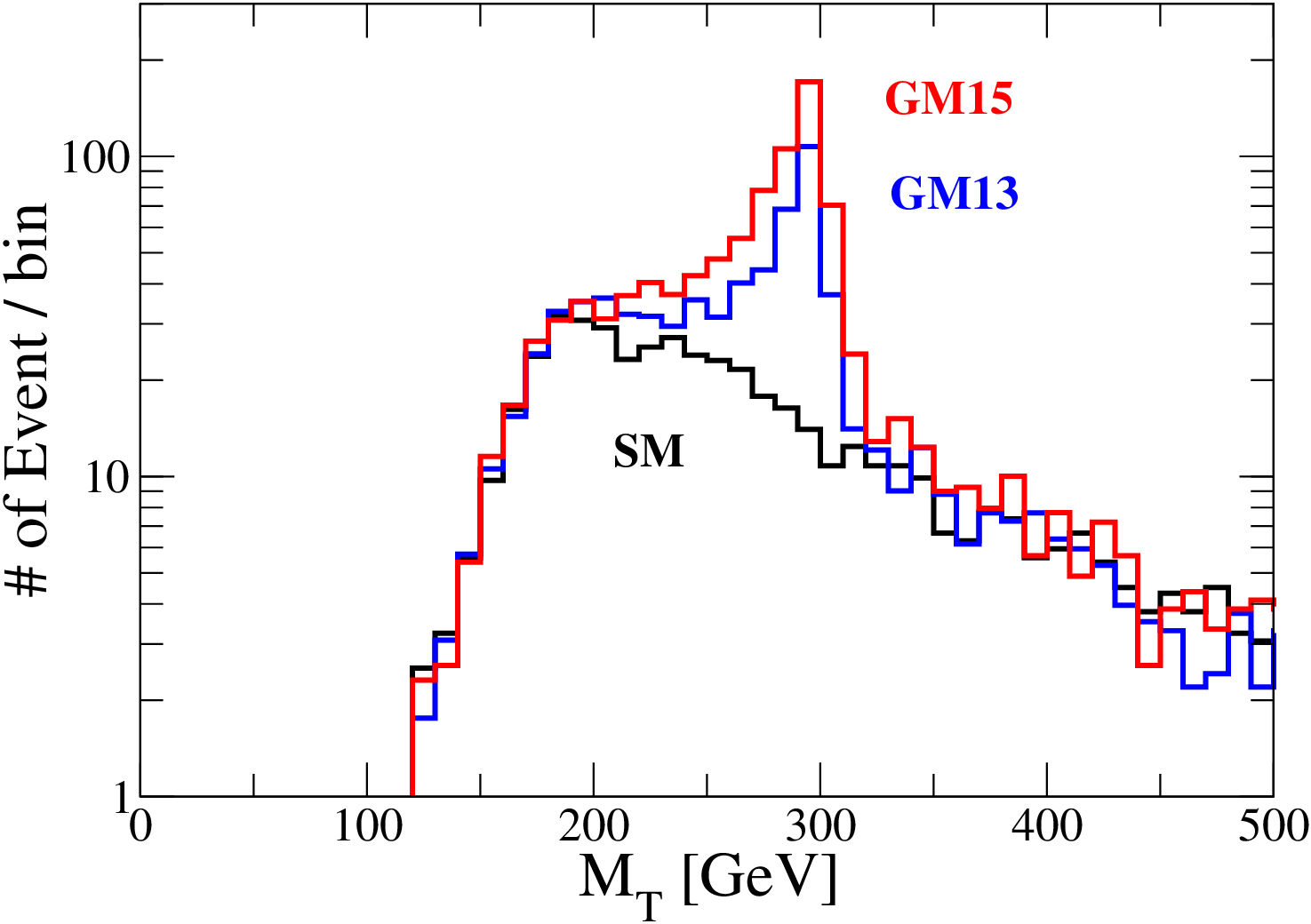}
\caption{Same as Fig.~\ref{distr_3l_8}, but for the collision energy of 14 TeV and 
an integrated luminosity of 100 fb$^{-1}$.} 
\label{distr_3l_14}
\end{center}
\end{figure}

Finally, we show the distributions of the $pp\to jj\ell^+\ell^+\ell^- E_T\hspace{-4.3mm}/\hspace{2mm}$ process after imposing the $\Delta\eta^{jj}$ cut in Figs.~\ref{distr_3l_8} and \ref{distr_3l_14}.  
Although there are three different combinations for the invariant mass distributions of two charged leptons, 
we only consider the $\ell^+\ell^+$ system to avoid the combinatorics issue for the $\ell^+\ell^-$ system. 
Again, we see in Fig.~\ref{distr_3l_8} significant excesses of the GM model over the SM background in 
the invariant mass distribution of the $\ell^+\ell^+$ system.  The transverse mass distribution of $\ell^+\ell^+\ell^- E_T\hspace{-4.3mm}/\hspace{2mm}$ system also features in a Jacobian-like peak with an edge at around 300 GeV, the mass of $H_5$.  This is due to the $H_5^+$ contribution to the $W^+Z\to W^+Z$ scattering.

A few remarks are now in order.
In the above analysis, we have not imposed any cut on the missing transverse energy, which is indispensable in real experimental analyses to remove reducible QCD backgrounds due to misidentifying soft jets as missing $E_T$.  If we add $E_T\hspace{-4.3mm}/\hspace{2mm}>100$ GeV to all the above-mentioned cuts, the cross sections are reduced to about 20$\%$.  However, the shapes of the distributions remain basically the same.
Obviously, if the masses of the $H_5$ states are varied, the shapes and edges of the corresponding bumps in the distributions will shift.
In addition, we have only evaluated the cross sections at the leading order.  To our knowledge, there is no full calculations of QCD corrections to all the vector boson scattering processes.  However, QCD corrections to the Higgs production via vector boson fusion processes have been worked out in Refs.~\cite{VBF_NLO1,VBF_NLO2,VBF_NLO3} at next-to-leading order (NLO) and subsequently at next-to-next-to-leading order (NNLO) in Refs.~\cite{VBF_NNLO1,VBF_NNLO2}.  According to Ref.~\cite{VBF_NNLO2}, the total cross section for the vector boson fusion production of a 125-GeV Higgs boson at NNLO differ from the SM is less than 10\%, less than the uncertainty due to choices of parton distribution functions.

Summarizing this section, we emphasize that an enhancement at and below 126 GeV in the lepton invariant mass distribution of the $\ell^+\ell^-E_T\hspace{-4.3mm}/\hspace{2mm}$ system would prefer models with $g_{hVV}$ couplings larger than the SM value, with the GM model being a well-defined example.
Moreover, it is a distinctive feature of the GM model to have an enhanced peak in the transverse mass distribution of the $\ell^+\ell^-E_T\hspace{-4.3mm}/\hspace{2mm}$ system with an edge at around 126 GeV and 
that of the $\ell^+\ell^+E_T\hspace{-4.3mm}/\hspace{2mm}$ and $\ell^+\ell^+\ell^-E_T\hspace{-4.3mm}/\hspace{2mm}$ 
systems at around the $H_5$ mass.

\section{Conclusions \label{sec:conclusions}}

Based on latest Higgs search data at the LHC, we have discussed the possibility of a larger-than-SM $g_{hVV}$ couplings between the SM-like Higgs boson and the weak gauge bosons.  Such a scenario can be readily realized in the Georgi-Machacek model or models with scalars of an appropriate representation.  We have performed $\chi^2$ fits to the signal strength data provided by the ATLAS and CMS Collaborations for various Higgs production channels at the LHC, showing that currently the $g_{hVV}$ couplings can be at most $\sim 1.3$ times the SM value at the 68$\%$ CL while consistent with several constraints, including the vacuum stability, the perturbative unitarity, the $Zb\bar{b}$ data and the electroweak precision data.  This result, along with the fact that the 126-GeV Higgs boson has largely unitarized the longitudinal weak boson scattering amplitudes, has led us to consider the scatterings in the multi-hundred GeV regime.
We selected two representative scenarios, GM 13 and GM15, to perform a simulation study for three types of processes involving the weak boson scatterings and compare them with the SM background.  The expected numbers of events after the basic and forward-jet tagging cuts for the LHC running at 8 and 14 TeV are given in Table~\ref{cross}.
In particular, we have shown that features due to the enhanced $g_{hVV}$ couplings can be easily identified by Jacobian-like peaks in the transverse mass distributions of the $\ell^+\ell^- E_T\hspace{-4.3mm}/\hspace{2mm}$, 
$\ell^\pm\ell^\pm E_T\hspace{-4.3mm}/\hspace{2mm}$ and $\ell^\pm\ell^\pm\ell^\mp E_T\hspace{-4.3mm}/\hspace{2mm}$ systems, with the edges signifying the masses of the Higgs bosons with the dominant contributions.  
Therefore, measuring such scattering events is not only important in determining whether the $g_{hVV}$ couplings are stronger than SM expectation, but also useful to discover additional Higgs bosons in the sub-TeV regime.

\section*{Acknowledgments}

We thank C.-M.~Kuo for useful experimental information.  C.-W.~C would like to thank the hospitality of the High Energy Theory Group at Rutgers University during his visit while this work is being finished.  This research was supported in part by the National Science Council of R.~O.~C. under Grant Nos.~NSC-100-2628-M-008-003-MY4 and NSC-101-2811-M-008-014.

\end{document}